\documentclass[aps]{revtex4}
\usepackage{graphicx}
\begin{document}
\title{The sum rule for the polarized structure function $g_2$
corresponding to the moment at $n=0$}
\author{Susumu Koretune and Hirofumi Kurokawa}
\affiliation{Department of Physics, Shimane University,
Matsue,Shimane,690-8504,Japan}
\begin{abstract}In the small $Q^2$ region, the sum rule for
the polarized structure function $g_2$ 
corresponding to the moment at $n=0$ is derived. 
This sum rule shows that there is a tight connection among the resonances,
the elastic and the continuum in the $g_2$. Further, the Born term
contribution in this sum rule is proportional to $Q^2$ and very small compared with that in the
corresponding sum rule for the polarized structure function 
$g_1$. However, the Born term contribution divided by $Q^2/2$ which 
also appears in the Schwinger sum rule 
for the $g_2$ corresponding to the moment at $n=1$ has a very similar
behavior with that in the sum rule for the $g_1$ 
corresponding to the moment at $n=0$.
\end{abstract}
\maketitle
The polarized structure functions $g_1$ and $g_2$ at low energy in the small
$Q^2$ region attract great interest recently.  The $\Delta (1232)$ gives
the large negative contribution in this
region and it can explain the sign difference between  Ellis-Jaffe
sum rule\cite{Ellis} and Gerasimov-Drell-Hearn sum rule\cite{Drell,Ger}. This $\Delta
(1232)$ contribution also invalidate a naive application of Bloom-Gilman
duality to the small $Q^2$ region\cite{BG}. Now, in this
region, we also have the continuum contribution and the large elastic
contribution. 
Recently, the sum rule for the $g_1$ in the small $Q^2$ region has been
derived, and it has been shown that there exists the 
tight connection among the resonances,the elastic
and the continuum in the $g_1$\cite{kore1,kore2}. In this paper we show that a similar sum rule exists
for the $g_2$.\\

According to Ref.\cite{DJT}, fixed-mass sum rules based on the canonical
quantization on the null-plane gives us
\begin{equation}
 \int_{0}^{1}\frac{dx}{x}g_1^{[ab]}(x,Q^2)=-\frac{1}{16} f_{abc}\int_{-\infty}^{\infty}
d\alpha \epsilon (\alpha )[A_c^5(\alpha ,0)+\alpha \bar{A}_c^5(\alpha ,0)],
\end{equation}
and
\begin{equation}
 \int_{0}^{1}\frac{dx}{x}g_2^{[ab]}(x,Q^2)=\frac{1}{16} f_{abc}\int_{-\infty}^{\infty}
d\alpha\epsilon (\alpha ) \alpha \bar{A}_c^5(\alpha ,0),
\end{equation}
where $x=Q^2/2\nu$ and $\nu =p\cdot q$, and $A_c^{5\beta}(x|0)$($x$ in this
expression is the space-time variable) is the anti-symmetric bilocal current, and its
matrix element is defined as
\begin{equation}
 <p,s|A_c^{5\beta}(x|0)|p,s>_c=s^{\mu}A_c^5(p\cdot x,x^2)+p^{\mu}(x\cdot
 s)\bar{A}_c^5(p\cdot x,x^2)+x^{\mu}(x\cdot s)\tilde{A}_c^5(p\cdot x,x^2).
\end{equation}
Similar sum rules can be derived from the current anti-commutation
relation on the null-plane\cite{kore2}. These sum rules are for the symmetric
combination under the interchange of superscript $a$ and $b$.
The basic difference between the sum rules based on the current commutation
relation and the current anti-commutation relation is
that the former ones are based on the operator relation while the latter
ones are based on the connected matrix element between the one particle
stable hadron. In this sense, the former sum rules
are more general than the latter ones. However, in the latter case,
the sum rules are directly applied to the structure functions in the
electroproduction, while in the former case, it is for the isovector
photon. The sum rules for the $g_1$ are given in Refs.\cite{kore1,kore2}
based on the fact that the right-hand side of Eq.(1) is $Q^2$ independent. 
The sum rule for the $g_2$ can be derived by the same kind of reasoning that 
the right-hand side of Eq.(2) is $Q^2$ independent as
\begin{equation}
 \int_{0}^{1}\frac{dx}{x}g_2^{ab}(x,Q^2)=\int_{0}^{1}\frac{dx}{x}g_2^{ab}(x,Q^2_0),
\end{equation}
where the superscript $ab$ is kept. In the current commutator case, it
takes the ones corresponding to the charged photon 
as in Ref.\cite{kore1}, and in the current anti-commutator case, 
it takes the ones corresponding to the usual electromagnetic current as
in Ref.\cite{kore2}.
Now since we have
\begin{eqnarray}
\triangle\sigma^{ab}(\nu ,Q^2)&=&\sigma^{ab}_{3/2}(\nu ,Q^2) - \sigma^{ab}_{1/2}(\nu ,Q^2)\\\nonumber
&=&-\frac{8\pi^2\alpha_{em}}{K}\left(\frac{g_1^{ab}(x,Q^2)}{\nu} - 
\frac{m_N^2Q^2g_2^{ab}(x,Q^2)}{\nu^3}\right),
\end{eqnarray}
where $\displaystyle{K=(1 - \frac{Q^2}{2\nu})}$, we have the following 
relation at $Q^2=0$
\begin{equation}
\frac{g_1^{ab}(x,0)}{\nu}=-\frac{1}{8\pi^2\alpha_{em}}
\triangle\sigma^{ab}(\nu ,0) .
\end{equation}
Thus the method to use the photo-reaction as the
regularization point can not be applied directly to the $g_2$.
Though we can take one particular reaction at small $Q^2$ as a
regularization point, the
relation with the real photon reaction is interesting in itself,
since the real and the virtual photon is essentially different.
Further, if we can derive a similar sum rule as the $g_1$,
we can consider the $g_1$ and the $g_2$ at the same footing.
Now if we differentiate Eq.(5) by $Q^2$ and take the limit $Q^2\to 0$, 
we obtain the relation
\begin{equation}
\frac{g_2^{ab}(x,0)}{\nu} = \frac{g_1^{ab}(x,0)}{2m_N^2}
+\left. \frac{\nu}{m_N^2}\frac{\partial g_1^{ab}(x,Q^2)}{\partial Q^2}\right|_{Q^2=0}
+\left.
  \frac{\nu^2}{8\pi^2m_N^2\alpha_{em}}\frac{\partial\triangle\sigma^{ab}(\nu, Q^2)}{\partial
Q^2}\right|_{Q^2=0}.
\end{equation}
All the quantities on the right-hand side are experimentally measurable.
Hence we can relate $g_2^{ab}(x,0)/\nu$ to the experimentally measurable
quantity. Then, by setting $Q^2=0$ on the right-hand side of Eq.(4),
we can rewrite the sum rule (4) by the same method as in the sum rule for the $g_1^{ab}$\cite{kore2}.
We first separate the Born term contribution and then cut off
the integral of the continuum part at some value in $E$ where $E$ is defined in the
laboratory frame as $\nu =p\cdot q= m_NE$. We denote this cutoff value as
$E_c$. Then we define the threshold value for the continuum as $E_0$,
and $E_0(Q)=E_0 + Q^2/2m_N, E_c(Q)=E_c + Q^2/2m_N, x_c(Q)=Q^2/(2m_NE_c(Q))$,
and take $E_c=2$(GeV).
In this way, we obtain the sum rule 
\begin{equation}
\int_{x_c(Q)}^1\frac{dx}{x}g_2^{ab}(x,Q^2) =
B_2^{ab}(Q^2) +\int_{E_0}^{E_c}\frac{dE}{E}g_2^{ab}(x,0) +
K_2^{ab}(E_c,Q^2),
\end{equation}
where $B_2^{ab}(Q^2)$ is the Born term at $Q^2=0$ minus the Born term at
$Q^2$ and $K_2^{ab}(E_c,Q^2)$ is given as
\begin{equation}
K_2^{ab}(E_c,Q^2) = \int_{E_c}^{\infty}\frac{dE}{E}g_2^{ab}(x,0) -
\int_{E_c(Q)}^{\infty}\frac{dE}{E}g_2^{ab}(x,Q^2),
\end{equation}
and the quantities on the right-hand
side in Eq.(7) is substituted for $g_2^{ab}(x,0)/E$ in Eqs.(8) and (9).
Further, through the regularization of the sum rule explained in
Ref.\cite{kore2}, the integral in Eq.(9) is taken after the subtraction
of the high energy behavior. Note that the integral on the left hand
side of Eq.(8) is restricted below $x_0(Q)=Q^2/2m_NE_0(Q)$ since the
Born term is separated out, where $E_0(Q)$
is determined by the threshold of the pion electroproduction as
$2m_NE_0(Q)=(m_N+m_{\pi})^2-m_N^2+Q^2$.\\

Now, in case of the proton target,the sum rule 
for the current commutation relation with 
$a=(1+i2)/\sqrt{2},b=a^{\dagger}$ is given by taking $g_2^{ab}(x,Q^2)$
and $B_2^{ab}(Q^2)$ which we denote $g_2^{+-}(x,Q^2)$ and $B_2^{+-}(Q^2)$
respectively as
\begin{equation}
g_2^{+-}(x,Q^2)=2g_2^{1/2}(x,Q^2)-g_2^{3/2}(x,Q^2),
\end{equation}
where the superscript $1/2$ or $3/2$ 
means the quantity in the reaction\\
(isovector photon) + (proton) $\to$ (states of isospin I) where
$I=1/2,3/2$, and  
\begin{equation}
B_2^{+-}(Q^2) = \frac{Q^2}{16m_p^2}\frac{1}{1+\frac{Q^2}{4m_p^2}}
G_M^+(Q^2)(G_M^+(Q^2)-G_E^+(Q^2)),
\end{equation}
where
\begin{eqnarray}
G_E^+(Q^2) &=& G_E^p(Q^2) - G_E^n(Q^2), \\\nonumber
G_M^+(Q^2) &=& G_M^p(Q^2) - G_M^n(Q^2),
\end{eqnarray}
and Sachs form factors $G_E^p(Q^2),G_M^p(Q^2)$ are normalized
as $G_E^p(0)=1,G_M^p(0)=\mu_p=2.793$.
It should be noted that the Born term contribution is proportional to $Q^2$, 
and hence its contribution is zero at $Q^2=0$. Further,we denote
$K_2^{ab}(E_c,Q^2)$ as $K_2^{+-}(E_c,Q^2)$. \\

In case of the current anti-commutation relation for the proton target,
we get the sum rules for the structure function in the
electroproduction, hence we denote $g_2^{ab}(x,Q^2)$ and $B_2^{ab}(Q^2)$ in
this case as $g_2^{ep}(x,Q^2)$
and $B_2^{ep}(Q^2)$ respectively. Further,we denote
$K_2^{ab}(E_c,Q^2)$ in this case as $K_2^{ep}(E_c,Q^2)$. The explicit
form of the Born term 
contribution is
\begin{equation}
B_2^{ep}(Q^2) =
 \frac{Q^2}{8m_p^2}\frac{1}{1+\frac{Q^2}{4m_p^2}}G_M^p(Q^2)(G_M^p(Q^2)-G_E^p(Q^2))
\end{equation}
Combined with a similar sum rule for the $g_1^{ep}$ in the previous
paper\cite{kore2} given as
\begin{equation}
\int_{x_c(Q)}^1\frac{dx}{x}g_1^{ep}(x,Q^2)  =
B_1^{ep}(Q^2)-\frac{m_p}{8\pi^2\alpha_{em}}\int_{E_0}^{E_c}dE\{\sigma_{3/2}^{\gamma p}-\sigma_{1/2}^{\gamma p}\} +
K_1^{ep}(E_c,Q^2),
\end{equation}
where $B_1^{ep}(Q^2)$ is given as
\begin{eqnarray}
B_1^{ep}(Q^2)&=&\frac{1}{2}\left\{F_1^p(0)[F_1^p(0)+F_2^p(0)]-
F_1^p(Q^2)[F_1^p(Q^2)+F_2^p(Q)]\right\}            \\\nonumber
&=&\frac{1}{2}\left\{\mu_p
	       -\frac{1}{1+\frac{Q^2}{4m_p^2}}[G_M^p(Q^2)(G_E^p(Q^2)
+\frac{Q^2}{4m^2}G_M^p(Q^2))]\right\} ,
\end{eqnarray}
and $K_1^{ep}(E_c,Q^2)$ as
\begin{equation}
K_1^{ep}(E_c,Q^2)=\frac{m_p}{8\pi^2\alpha_{em}}\int_{E_c}^{\infty}dE
\{\sigma_{1/2}^{\gamma p}-\sigma_{3/2}^{\gamma p}\} -
\int_{E_c(Q)}^{\infty}\frac{dE}{E}g_1^{ab}(x,Q^2),
\end{equation}
we obtain the sum rule for the $(g_1^{ep} + g_2^{ep})$ as
\begin{equation}
\int_{x_c(Q)}^1\frac{dx}{x}(g_1^{ep}(x,Q^2) + g_2^{ep}(x,Q^2)) =
B_1^{ep}(Q^2)+ B_2^{ep}(Q^2)+\int_{E_0}^{E_c}\frac{dE}{E}(g_1^{ep}(x,0)+g_2^{ep}(x,0)) +
K_1^{ep}(E_c,Q^2)+K_2^{ep}(E_c,Q^2).
\end{equation}
The explicit form of the Born term is
\begin{equation}
B_1^{ep}(Q^2)+ B_2^{ep}(Q^2) = \frac{1}{2}(\mu_p -G_M^p(Q^2)G_E^p(Q^2)).
\end{equation}
The magnitude of the Born term contributions in the moment at $n=0$ for the
$g_1^{ep}$ and the $(g_1^{ep}+g_2^{ep})$ are very similar,
but that of the $g_2^{ep}$ is very small compared with these
since it is proportional to $Q^2$. However, if 
this Born term is divided by $Q^2/2$, it has a
finite limit as $Q^2\to 0$, and has an interesting behavior. 
These quantities are the ones which appear in the
Schwinger sum rule for the $g_2^{ep}$ given as\cite{Schw}
\begin{equation}
\frac{-1}{4m_p^2+Q^2}G_M^p(Q^2)(G_M^p(Q^2)-G_E^p(Q^2))
+\int_{\nu_0(Q)}^{\infty}d\nu G_2^{ep}(\nu ,Q^2) = 0,
\end{equation}
where we separate the Born term in this sum rule.
At large $Q^2$, because of the Burkhart-Cottingham(BC) sum
rule\cite{BC} for the inelastic reaction, we have the relation
\begin{equation}
I(Q^2)=\int_{\nu_0(Q)}^{\infty}d\nu G_2^{ep}(\nu ,Q^2)=\frac{2}{Q^2}
\int_{0}^{1}dxg_2^{ep}(x,Q^2)=0.
\end{equation}
Thus we can consider the main contribution in the continuum
part in the Schwinger sum rule (19) comes from a relatively low
energy region. Therefore, in the sum rule given as
\begin{equation}
\int_{\nu_0(Q)}^{\infty}d\nu G_2^{ep}(\nu ,Q^2) -
 \int_{\nu_0}^{\infty}d\nu G_2^{ep}(\nu ,0)
= B_S^{ep}(Q^2),
\end{equation}
where 
\begin{equation}
B_S^{ep}(Q^2) = \frac{1}{4m_p^2+Q^2}G_M^p(Q^2)(G_M^p(Q^2)-G_E^p(Q^2))
-\frac{\mu_p(\mu_p-1)}{4m_p^2},
\end{equation}
the main contribution on the left hand side comes from the
low $Q^2$ region. Since the Born term contribution $B_S(Q^2)$ changes
rapidly in this region, the left hand side of the sum rule also
changes rapidly. Since we have the relation $\nu = Q^2/2$ at the elastic
point,  $B_S^{ep}(Q^2)$ is related to $B_2^{ep}(Q^2)$ as
\begin{equation}
B_S^{ep}(Q^2) = \frac{2}{Q^2}B_2^{ep}(Q^2) - \left. \left\{
\frac{2}{Q^2}B_2^{ep}(Q^2)\right\}\right|_{Q^2=0}.
\end{equation}
Now the contribution to the quantity
\begin{equation}
\int_{x_c(Q)}^1\frac{dx}{x}g_2^{ep}(x,Q^2) - 
  \int_{x_c}^1\frac{dx}{x}g_2^{ep}(x,0)
\end{equation}
in the sum rule (8) comes from the low energy region
and we can expect it roughly given by $B_2^{ep}(Q^2)$. 
Thus the sum rule (8) and the Schwinger sum rule gives
us the same picture that the rapid behavior of the elastic
is compensated by the rapid behavior of the resonance and the continuum.
Now if we plot the Born term contributions $B_1^{ep}(Q^2),B_1^{ep}+B_2^{ep}(Q^2),$
and $-B_S^{ep}(Q^2)$, we find that these three functions behave very
similarly. As is shown in Figure, the difference between $B_1^{ep}(Q^2)$
and $-B_S^{ep}(Q^2)$ is very small and moreover the difference is almost
constant.\\
\begin{figure}
\includegraphics{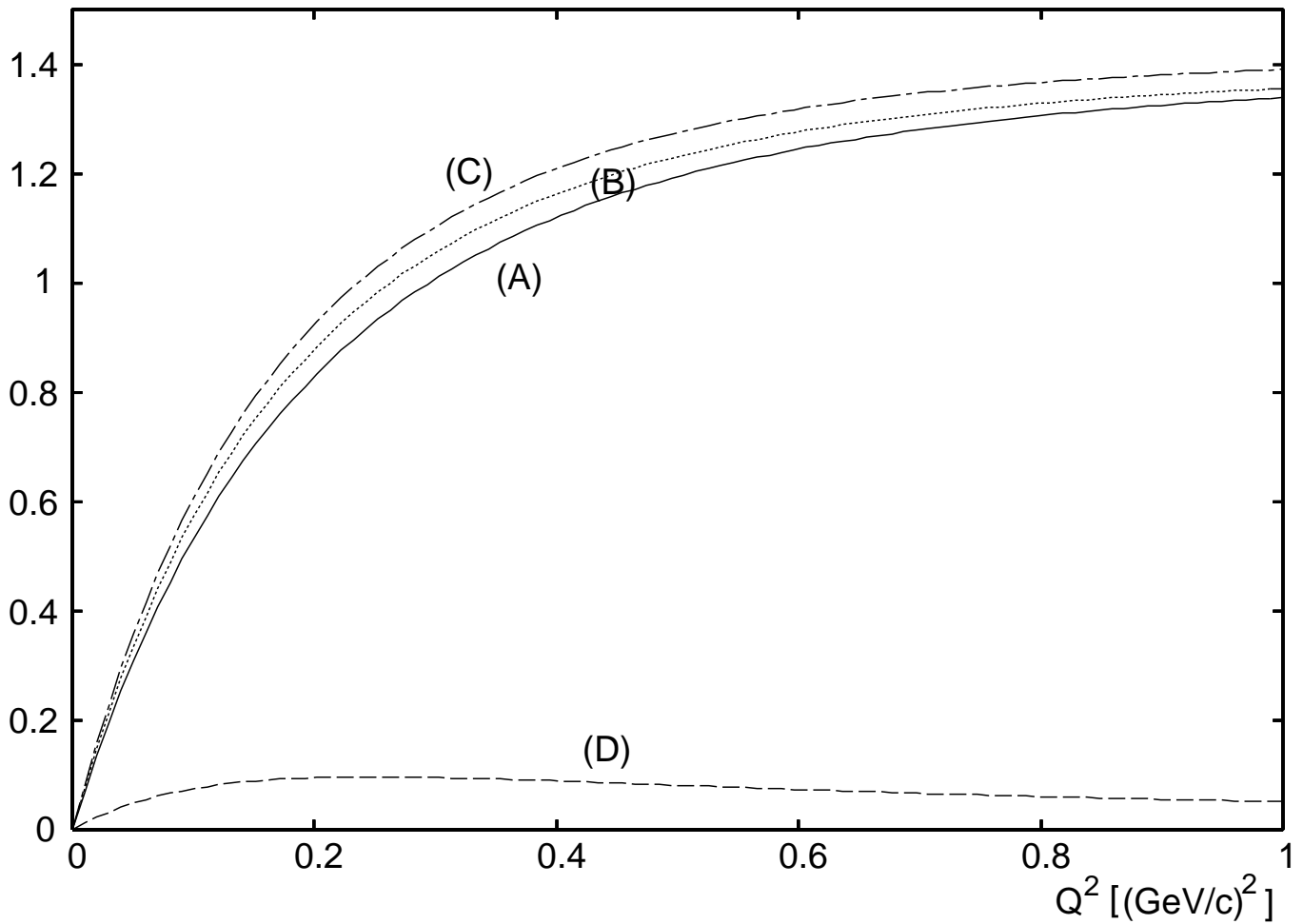}
\caption{\label{fig}The various Born term contributions. (A) is the 
$B_1^{ep}(Q^2)$ given in Eq.(15),
(B) is the $B_1^{ep}(Q^2)+B_2^{ep}(Q^2)$ given in
 Eq.(18), and (C) is the $-B_S^{ep}(Q^2)$ given in Eq.(22).
(D) is the difference between $B_1^{ep}(Q^2)$ and $(-B_S^{ep}(Q^2))$.}
\end{figure}
Though the moments which give $B_S^{ep}(Q^2)$ and $B_1^{ep}(Q^2)$
are different, we see that the behavior of the integral of
$\{-2g_2^{ep}(x,Q^2)/Q^2+(2g_2^{ep}(x,Q^2)/Q^2)|_{Q^2=0}\}$ and that of
$\{g_1^{ep}(x,Q^2)/x - (g_1^{ep}(x,Q^2)/x)|_{Q^2=0}\}$ in the small $Q^2$
region is very similar. Since the latter is related to the sign change
of the generalized Gerasimov-Drell-Hearn sum, this fact may suggest that
the $g_2^{ep}$ is related to this phenomena\cite{ST}. However, in our approach,
we have no direct relation between the $g_1^{ep}$ and the $g_2^{ep}$.

Concerned with this, we should point out that the seeming relation
between the $g_1^{ab}$ and the $g_2^{ab}$ in Eq.(7). This relation does not mean that
the $g_1^{ab}$ is related to the $g_2^{ab}$. However, if we substitute the experimental values
for the quantities on the right-hand side of Eq.(7), 
the $g_2^{ab}(x,0)$ determined by this relation depends on these values.
In this sense, the dependence on the $g_1^{ab}$ enters. Since the relation (7)
depends on the $Q^2$ dependence of $K$, and since we can extract
an experimental value even if we modify this flux factor, we can have
another sum rule by changing this factor. For example,
let us take $\bar{K}$ as
\begin{equation}
\bar{K} = 1 + b\cdot\frac{m_N^2Q^2}{\nu^2},
\end{equation}
where $b$ is an arbitrary dimension-less number, and $\bar{K}$
must be 1 at $Q^2 = 0$ since $\triangle\bar{\sigma}^{ab}(\nu ,Q^2)$
defined through $\bar{K}$
\begin{eqnarray}
\triangle\bar{\sigma}^{ab}(\nu ,Q^2)&=&\bar{\sigma}^{ab}_{3/2}(\nu ,Q^2)
 -  \bar{\sigma}^{ab}_{1/2}(\nu ,Q^2)\\\nonumber
&=&-\frac{8\pi^2\alpha_{em}}{\bar{K}}\left(\frac{g_1^{ab}(x,Q^2)}{\nu} - 
\frac{m_N^2Q^2g_2^{ab}(x,Q^2)}{\nu^3}\right),
\end{eqnarray}
must becomes quantity in the photoproduction. Then Eq.(7) changes as
\begin{equation}
\frac{g_2^{ab}(x,0)}{\nu} = -\frac{bg_1^{ab}(x,0)}{\nu}
+\left. \frac{\nu}{m_N^2}\frac{\partial g_1^{ab}(x,Q^2)}{\partial Q^2}\right|_{Q^2=0}
+\left.
  \frac{\nu^2}{8\pi^2m_N^2\alpha_{em}}\frac{\partial\triangle\bar{\sigma}^{ab}(\nu, Q^2)}{\partial
Q^2}\right|_{Q^2=0}.
\end{equation}
In this case, $g_1^{ab}(x,0)/\nu$ appears instead of $g_1^{ab}(x,0)$. Then
by using the sum rule (8) and the sum rule for the $g_1^{ab}$ given in Eq.(14) 
we obtain the sum rule for the $(bg_1^{ab}+g_2^{ab})$. This sum rule looks different from
the sum rule (14) even if we take $b=1$. This seeming difference
is the artifact of the difference of the definition of
$\triangle\bar{\sigma}^{ab}(\nu, Q^2)$
and $\triangle\sigma^{ab}(\nu, Q^2)$. Then we see that how we reach
the $Q^2=0$ point we have many different forms of the sum rule which
are essentially the same one.\\

In conclusion, in the small $Q^2$ region, we have derived the sum rule for
the polarized structure function $g_2$ corresponding to the moment at
$n=0$, which is similar to the corresponding sum rule for the $g_1$.
The $g_2$ at $Q^2=0$ is related to the experimentally measurable quantity, and it is
shown that the sum rule in appearance depends on how we reach the $Q^2=0$
point but that these seeming different sum rules are essentially the
same one. Then,independent of the $g_1$, we show that
there is a tight connection among the resonances,
the elastic and the continuum in the $g_2$. Since the Born term
contribution is proportional to $Q^2$ and very small compared with that in the
corresponding sum rule for the $g_1$, the change of the sum of 
the resonances and the continuum is small in this sum rule.
However, if we divide the Born term contribution in the sum rule for the $g_2$ 
by $Q^2/2$, which also appears in the Schwinger sum rule 
for the $g_2$ corresponding to the moment at $n=1$,the quantity obtained
has a very similar behavior with the Born term contribution 
in the sum rule for the $g_1$ corresponding to the moment at $n=0$.  
Whether this similarity is a mere happening or has a deep physical
meaning is not yet clear and needs a further study.

\end{document}